\documentclass[manuscript]{aastex}
\usepackage{subfigure}
\usepackage{amsmath}

\slugcomment{Draft, revised 3, \today}

\shorttitle{New Suns in the Cosmos?}
\shortauthors{D. B. de Freitas et al.}

\begin{document}

\title{New Suns in the Cosmos?}

\author{D. B. de Freitas, I. C. Le\~ao, C. E. F. Lopes, F. Paz-Chinchon, B. L. Canto Martins, S. Alves, J. R. De Medeiros}
\affil{Departamento de F\'{\i}sica,
    Universidade Federal do Rio
    Grande do Norte, 59072-970
    Natal, RN, Brazil}
\and
\author{M. Catelan}
\affil{Pontificia Universidad Cat\'olica de Chile, Departamento de Astronom\'{i}a y Astrof\'{i}sica, 
       Av. Vicu\~na Mackenna 4860, 782-0436 Macul, Santiago, Chile}
\and 	   
\affil{The Milky Way Millennium Nucleus, Santiago, Chile}

\begin{abstract}
The present work reports on the discovery of three stars that we have identified to be rotating Sun-like stars, based on rotational modulation signatures inferred from light curves from the CoRoT mission's Public Archives. In our analysis, we performed an initial selection based on rotation period and position in the Period--$T_{\rm eff}$ diagram. This revealed that the stars CoRoT IDs 100746852, 102709980, and 105693572 provide potentially good matches to the Sun with similar rotation period. To refine our analysis, we applied a novel procedure, taking into account the fluctuations of the features associated to photometric modulation at different time intervals and the fractality traces that are present in the light curves of the Sun and of these ``New Sun'' candidates alike. In this sense, we computed the so-called Hurst exponent for the referred stars, for a sample of fourteen CoRoT stars with sub- and super-solar rotational periods, and for the Sun, itself, in its active and quiet phases. We found that the Hurst exponent can provide a strong discriminant of Sun-like behavior, going beyond what can be achieved with solely the rotation period itself. In particular, we find that CoRoT ID 105693572 is the star that most closely matches the solar rotation properties, as far as the latter's imprints on light curve behavior is concerned. The stars CoRoT IDs 100746852 and 102709980 have significant smaller Hurst exponents than the Sun, notwithstanding their similarity in rotation periods. 
\end{abstract}


\section{Introduction}\label{sec:intro}

The question whether the Sun is typical or atypical as compared with other stars is one of the most exciting topics in present-day science, which has been addressed by a large number of previous studies. Indeed, quoting Gustafsson (2008), {\em Is the Sun unique as a star~-- and, if so, why?} The question of the normality of the Sun has known a large upsurge along the past 15 years, with the discovery of many extra-solar planetary systems, demonstrating that the Sun is not unique as a planet host star.  In this context, the main question now is the extent to which the properties of the Sun and its planetary system can be considered as representative of the planetary systems found around other stars. For instance, the discovery of giant gaseous planets close to many solar-type stars raises an additional question \citep[e.g.,][]{mfea10}: Is the Solar System, with Jupiter and Saturn at considerable distances from the Sun, really normal in this respect, and if so, is this just circumstantial, or could it be linked to the Solar System's zone of habitability~-- and, as a consequence, to our own existence?

Using a procedure based in comparing solar to stellar properties from particular stellar samples, some studies suggest that the Sun is a typical star (e.g., Gustafsson 1998; Allende Prieto 2008), whereas others suggest that it is atypical (e.g., Gonzalez 1999a,b; Gonzalez et al. 2001). These studies compared essentially mass, age, chemical composition, differential rotation, granulation and turbulence, activity and binarity. Several of these properties could be related to the habitability of a planetary system. In particular, rotation plays a fundamental role in a star's formation and evolution, controlling, in particular, magnetic fields and stellar winds. Therefore, studies of the solar rotation in comparison with other stars can bring out relevant information, as far as the question of the solar normalcy is concerned. Indeed, Soderblom (1983) and Gray (1982) have suggested that the Sun does rotate normally for its age and effective temperature, a result corroborated by Robles et al. (2008). Let us recall that these conclusions are based on the measurements of projected rotational velocity $v\sin i$, which depends on the inclination angle $i$ of the stellar rotational axis to the line of sight~-- which is usually unknown. In this sense, the CoRoT (Baglin 2006) and Kepler (Koch et al. 2010) space missions, offering the possibility of measuring rotation periods for thousands of stars, open a new route towards the study of the normalcy of the Sun vis-\`{a}-vis other planetary systems in the Cosmos. 

Several activity indicators (e.g., spots, flares) contribute to the complex temporal dynamics of stellar rotation effects. This renders light curves (hereafter LCs) extremely inhomogeneous and nonstationary, due to irregular fluctuations that are brought about by these phenomena. This also indicates that there are flutuactions acting in different scales, thus suggesting that a fractal analysis can provide a method to investigated changes in the scaling properties. A large number of studies (e.g., Hurst, Black, \& Simaika 1965; Mandelbrot \& Wallis 1969a; Feder 1988; Ruzmainkun, Feynman, \& Robinson 1994; Komm 1995; Kilcik et al. 2009; Suyal, Prasad, \& Singh 2009) using solar time-series (sunspot number, sunspot area) have shown that the so-called Hurst exponent $H$ (Hurst 1951) provides an efficient and powerful statistical method to investigate nonstationary fluctuations in solar data. Here we report on the discovery of three stars that are ``New Sun'' candidates, at least from the rotational point of view, based on detailed analysis of the rotational signatures obtained from LCs collected by the CoRoT space mission. In our analysis we applied a novel procedure, taking into account not only the variability of the features associated to photometric modulation at different time intervals, but also the fractality traces that are present in the LCs of the Sun (active and quiet) and of the ``New Sun'' candidates. The {\em Letter} is organized as follows. In \S\ref{sec:data} we describe the data used in our study. In \S\ref{sec:method} we describe the methods used in our analaysis of these data. Finally, in \S\ref{sec:results} we provide our results and discuss their implications. 

\section{Data and candidate selection}\label{sec:data}
\begin{figure}[!htb]
     \begin{center}
        \subfigure[]{%
            \label{fig1a}
            \includegraphics[width=0.5\textwidth]{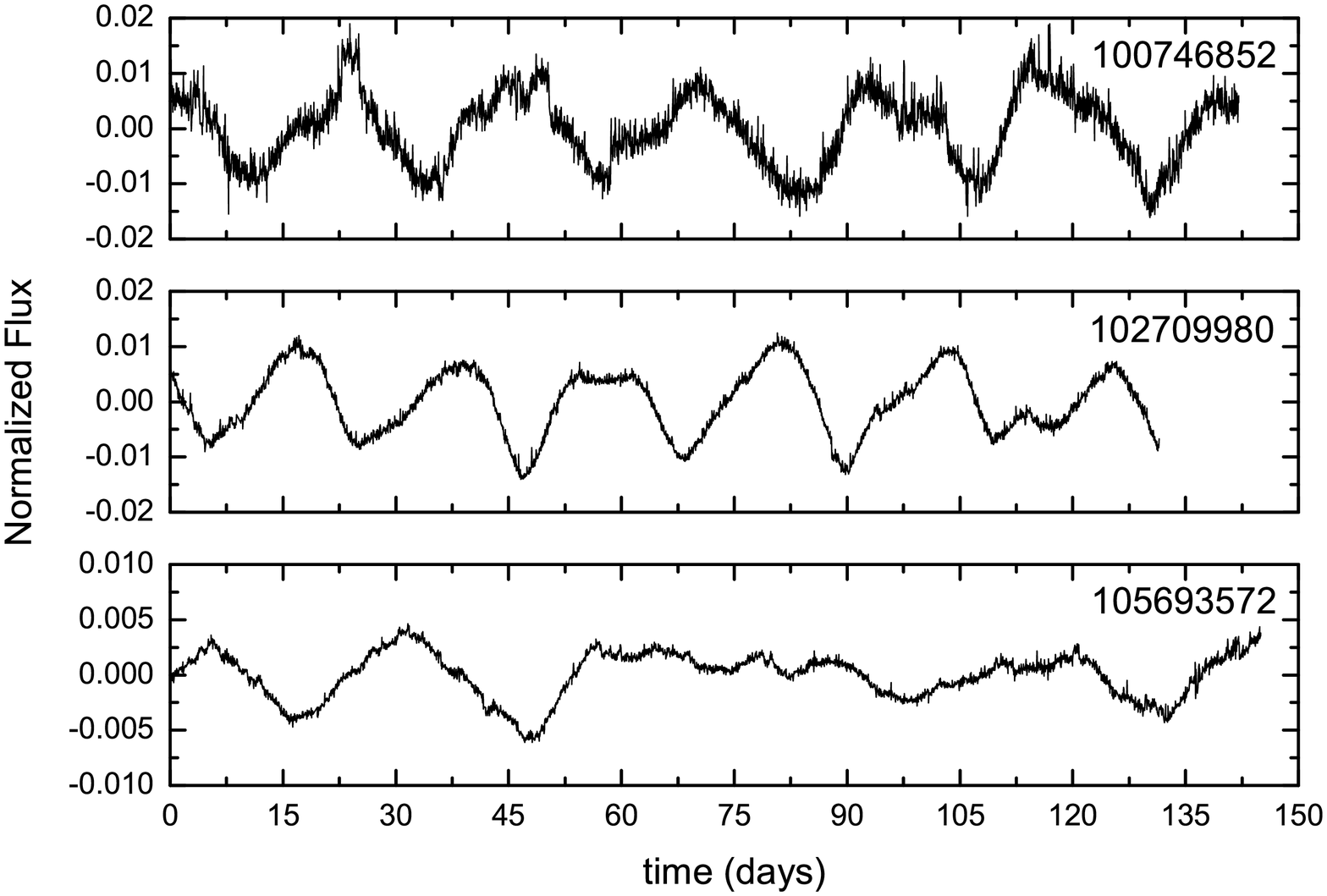}
        }%
        \subfigure[]{%
            \label{fig1b}
            \includegraphics[width=0.5\textwidth]{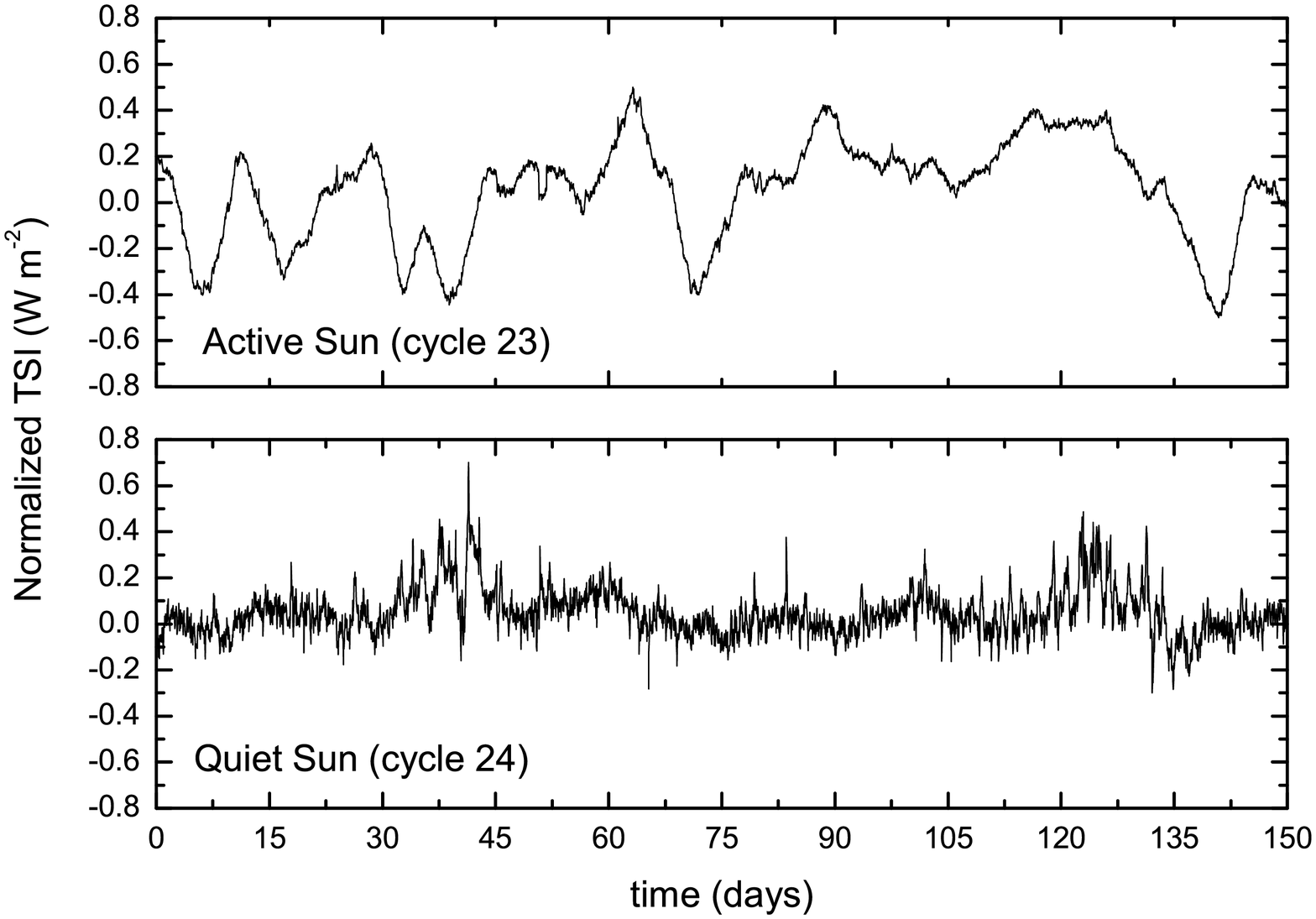}
        }\\
        \subfigure[]{%
           \label{fig1c}
           \includegraphics[width=0.5\textwidth]{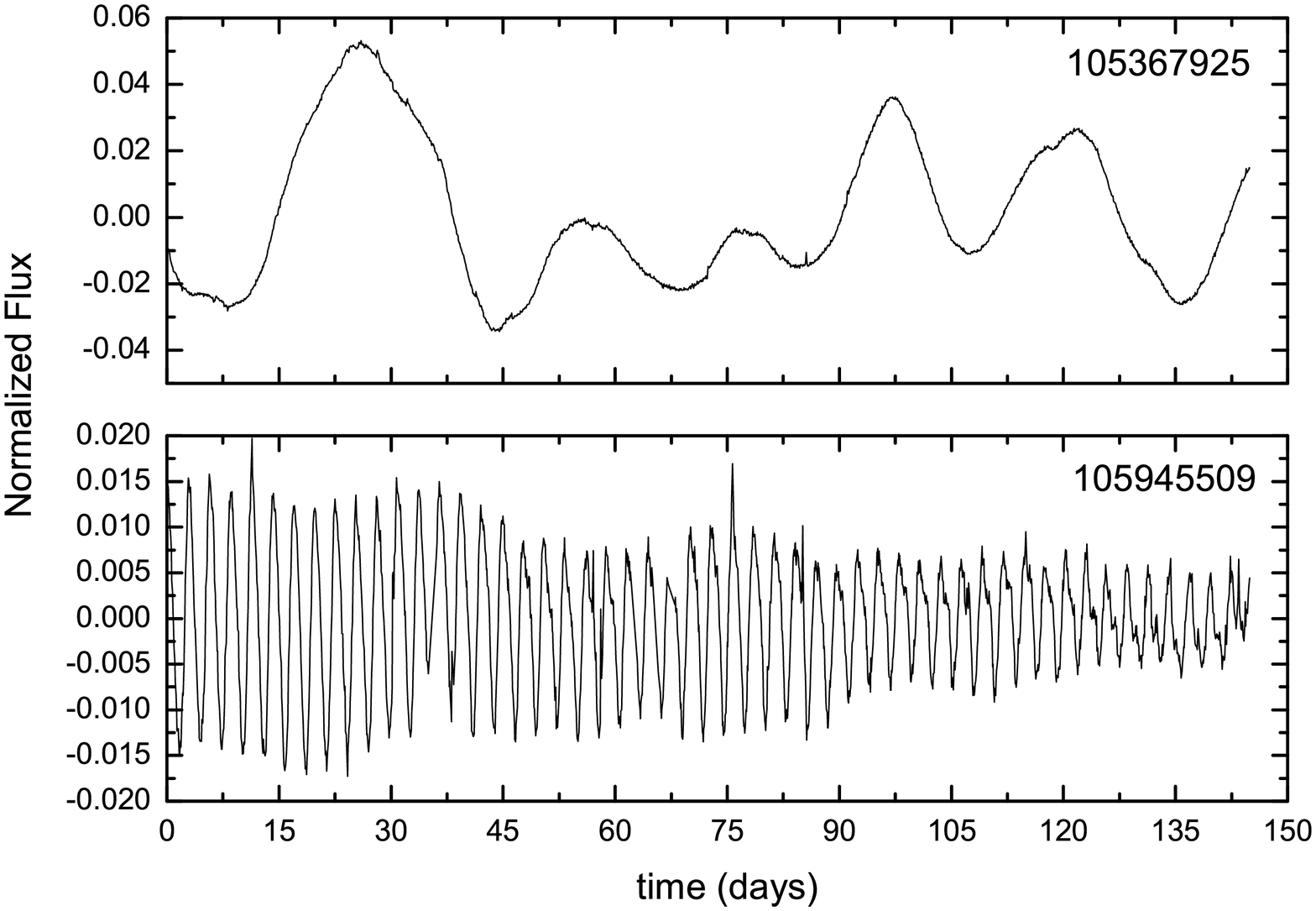}
   
        }%
       
    \end{center}
    \caption{%
        {\bf Figure (a):} CoRoT LC for the ``New Sun'' candidates} obtained as described in De Medeiros et al. (2013). {\bf Figure (b):} TSI time series, based on Virgo/SoHO measurements, for two solar phases: active ({\em upper panel}) and quiet ({\em lower panel}). {\bf Figure (c): An extract from non-Sun CoRoT LC.
     }%
   \label{fig1}
\end{figure}

De Medeiros et al. (2013) have produced a list of 4206 stars presenting unambiguous semi-sinusoidal variability signatures in their LCs, as obtained in the course of the CoRoT mission, compatible with rotational modulation. From that sample, three stars, CoRoT IDs 100746852, 102709980, and 105693572, present periods in the range $\approx 25 \pm 5$~days, i.e., matching closely the Sun's rotation period, which is 24.47~days at the equator, 33.5~days on the poles, and 26.09~days on average \citep{lanza2003,em12}. Physical parameters of the selected candidates, including their  computed rotation periods, are listed in Table~\ref{tab1}. Readers are referred to De Medeiros et al. (2013) for a description of LCs treatment. Nevertheless, let us briefly mention that possible outliers, discontinuities, and/or long-term trends present in the analyzed LCs were corrected following the prescriptions described by those authors, based on previous works on the subject (e.g., Mislis et al. 2010; Basri et al. 2011). Lomb-Scargle periodogram (Lomb 1976; Scargle 1982) was used to compute the solar rotation rates and CoRoT stars periods. For each LC, a periodogram was computed for periods with false alarm probability ${\rm FAP} < 0.01$ (i.e., significance level $> 99\%$). Error on the computed period was
estimated by considering a harmonic fit of the phase diagram, with four harmonics. The period range in which the harmonic model does not exceed 1-$\sigma$ of the data residual from the best fit, within the observation
time span, was assumed to be the error (see 7th column from Table~\ref{tab1}). The CoRoT LCs for the discovered stars are displayed in Figure~\ref{fig1a}. Originally, the CoRoT LCs typically have a cadence of
32~s for particular targets and 512~s for regular targets. The time span
of these LCs is 54--57~days for the Initial Run (IRa01) and 131, 142--152,
and 144~days for the long runs LRa01, LRc01 and LRc02, respectivelly. For homogeneity with solar time series, the cadence for all these LCs was fixed to one-hour (0.0417~day), corresponding to the cadence for Solar Data. For the determination of the stellar parameter $T_{\rm eff}$ readers are referred to Sarro et al. (2013). 

For comparison purposes, we also analyze the Total Solar Irradiance (TSI) time series (Lanza et al. 2003). From the publicly available TSI data obtained by the Virgo/SoHO team,\footnote{{\tt http://virgo.so.estec.esa.nl/}} which were taken on an hour-by-hour basis and have an internal precision of $2.0 \times 10^{-5}$, we selected the Sun in two phases: active (cycle 23) and quiet (cycle 24) as shown in Figure.~\ref{fig1b}. The time windows in each phase of solar activity were chosen because of their similarity to the CoROT data for the ``New Sun'' candidates, in terms of the smaller residual difference between the stars and the Sun LC's. In addition to the Sun itself, we also analyzed a subsample of fourteen stars observed by CoRoT, with rotation periods markedly longer or shorter than the solar values. The properties of these stars are summarized in Table~\ref{tab1}.  

\begin{figure}
\plotone{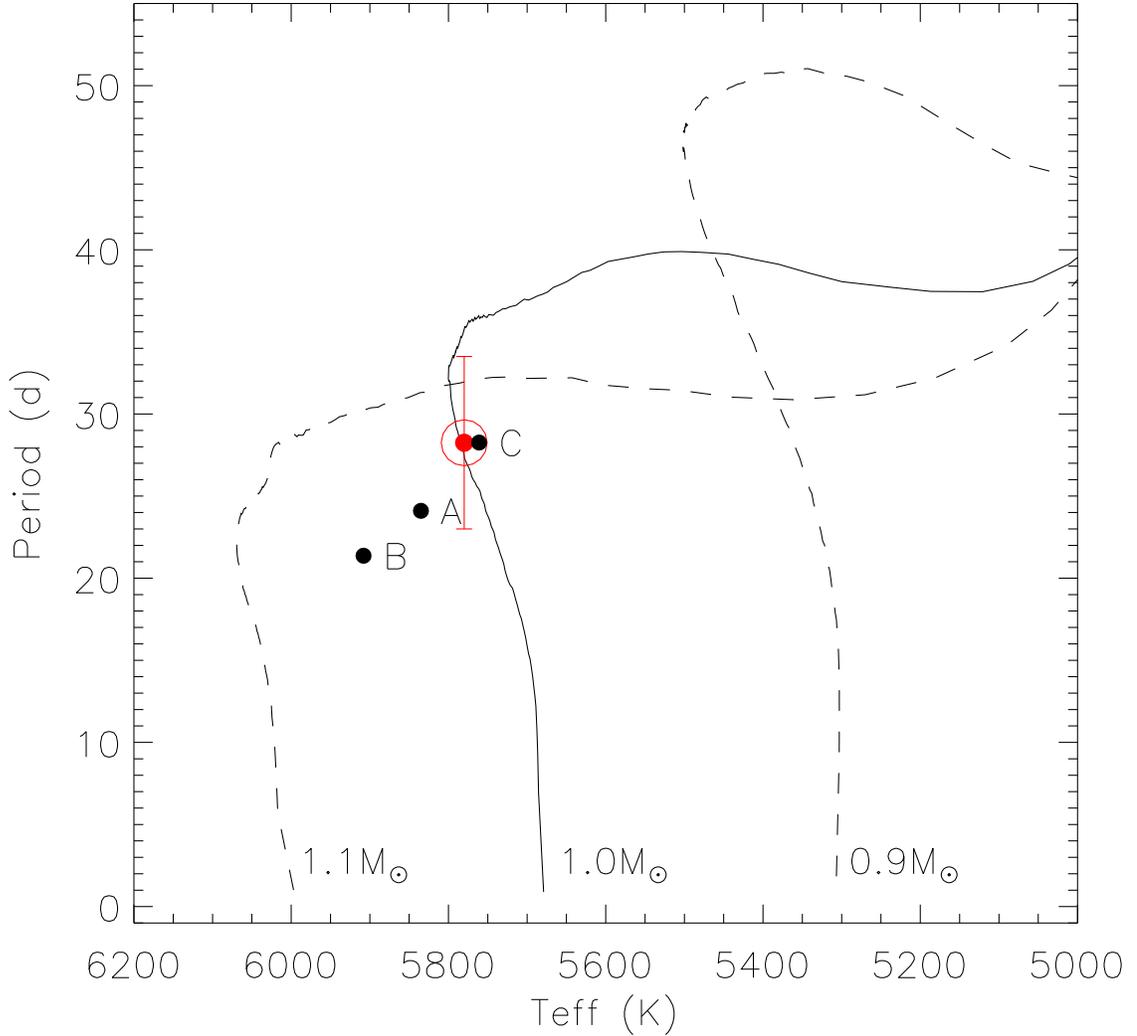}
\caption{Period versus $T_{\rm eff}$ diagram for our selection of ``New Sun'' candidates, with 
physical parameters taken from Sarro et al. (2013).
The typical errors are 3\% for the period and $\sim$400~K for $T_{\rm eff}$.
The targets are identified by the letters A to C, corresponding to stars with 
CoRoT IDs 100746852, 102709980, and 105693572, respectively.
Variability periods came from De Medeiros et al. (2013).
The Sun is illustrated by its usual symbol, with a bar representing its 
observed range in rotation periods (e.g., Lanza et al. 2003).
Solar-metallicity evolutionary tracks from Ekstr{\"o}m et al.~(2012)
for masses in the range between 0.9 and $1.1 \, M_\odot$ are overplotted on the data.}
\label{fig2}       
\end{figure}

The computed periods for stars CoRoT IDs 100746852, 102709980, and 105693572, as given in Table~\ref{tab1}, are very close to the solar values, pointing to a similarity in rotation between these stars and the Sun.
An additional aspect reinforcing such similarity comes from the position of the referred stars in the period versus $T_{\rm eff}$ diagram, as displayed in Figure~\ref{fig2}, in particular for the star CoRoT ID 105693572, with a 28-day rotation period. Solar-metallicity evolutionary tracks for 0.9, 1.0, and $1.1 \, M_\odot$ from Ekstr{\"o}m et al. (2012) are also overplotted on the data. For these tracks were, those authors considered the effects of rotation in a homogeneous way
for an average evolution of non-interacting stars, by accounting for both atomic diffusion and magnetic braking in low-mass star
models. Within the uncertainties given by Sarro et al. (2013) for $T_{\rm eff}$, the positions of these stars in the period versus $T_{\rm eff}$ diagram are closely similar to the Sun's.
Indeed, as underlined by these authors, the typical error on the temperature is relatively large ($\sim$400~K). 

\section{A new approach to rotational similarity}\label{sec:method}
In addition to the similarity in rotation periods and Hertzsprung-Russell Diagram (HRD) position between the three selected stars and the Sun, we are now in a position to also check, in unprecedented detail, how similar the actual morphology of their LCs really are to that of the Sun. In fact, stellar photometric time series or LCs reflect a mixture of complex processes, such as rotational modulation, oscillation, and magnetic activity, among others. Variations in the spatial and temporal distribution of small and large structures on the stellar photosphere can lead to complex variability signatures in the LC, going well beyond just the single datum provided by the period of rotation. In general, LC variability can be extremely irregular, inhomogeneous and (multi)fractal.

To analyse the behavior of variability and its fluctuation in different spatio-temporal scales, it is mandatory to perform one or more techniques of statistical analysis of the LC data. In general, a time series would be given by a deterministic function in the absence of noise. ``Unfortunately'', time series always incorporate some degree of stochastic noise, which can be characterized by a specific ``color'' (Carter \& Winn 2009). Fractals can be considered as a form of colored noise present in LCs defined by profile of power law frequency spectrum, characterized by a Power Spectral Density (PSD) $S(f)$ at frequency $f$ given by $S(f)=\frac{A}{f^{\alpha}}$, with $\alpha$ estimated from the determination of the negative slope of a linear trend (Pascual-Granado 2011). The $\alpha$ exponents, which allow a distinction between fractional Gaussian noise (fGn) and fractional Brownian motion (fBm) series (Eke et al. 2000, Delignieres et al. 2005), are related to Hurst exponent $\hat{H}$ by the relation $\alpha=2\hat{H}\pm1$, sign ``$+$''  for fBm and sign ``$-$'' for fGn (Gao et al. 2006). A special case is $\alpha=0$ noise, from which the PSD is flat at all frequencies indicating that the noise is uncorrelated. The increasing of $\alpha$ leads to longer-range correlation in time. Here, $\hat{H}$ represents the values estimated by PSD analysis and $H$ denotes the true exponent of the LC. 

According to Ivanov et al. (1999), homogeneous time series can be indexed by a single, global Hurst exponent $H$ (Hurst 1951). In contrast, complex time series can be decomposed in a wide spectrum characterized by different, local Hurst exponents. This exponent is a parameter that quantifies the persistent or anti-persistent (i.e., past trends tend to reverse in the future) behavior of a time series (Suyal et al. 2009; Kilcik et al. 2009). The value of the exponent $H$ distinguishes the behavior of the time series. When applied to fBm, if $H = 0.5$, the time series is purely random or Brown motion, normally distributed or has no memory. If $H > 0.5$, the time series is said to be persistent, that is, the data cover more ``distance'' than a random walk. Finally, for $H<0.5$, the time series is antipersistent. A particular case is $H=1$, which denotes periodic motions similar to sinusoidal variations. In this context, the Hurst exponent may also be used as a measure of complexity degree, with a smaller $H$ denoting a more complex system (see Yu and Chen 2000).

In order to estimate the Hurst exponent $H$ in LCs, we apply the well-known rescaled range ($R/S$) method proposed by Mandelbrot \& Wallis (1969b) and originally developed by Hurst (1965), following the procedure by Martinis et al. (2004), which computes te Hurst exponent using the values of the rescaled range $R/S$ over a box of $n$ elements as a power law given by $(R/S)=k \, n^{H}$, where $k$ is a constant. The procedure starts with two elements, $n=2$ and for each iteration one more element is added up to $n=N$ (i.e., the whole series), with $R/S$ calculated for the wider box. The slope of the least-square linear fit in a log-log plot gives the value of the Hurst exponent. In our study, we employ the Hurst exponent via $R/S$ method to classify the CoRoT LCs into three rotation regimes, namely sun-like rotation period (``New Suns''), sub- and super-sun rotational periods. This procedure pronounces clear differences among these regimes as indicated in Figures~\ref{fig3a} and \ref{fig3b}.

\section{Results and Discussions}\label{sec:results}

\begin{figure}[!htb]
     \begin{center}
        \subfigure[]{%
            \label{fig3a}
            \includegraphics[width=0.5\textwidth]{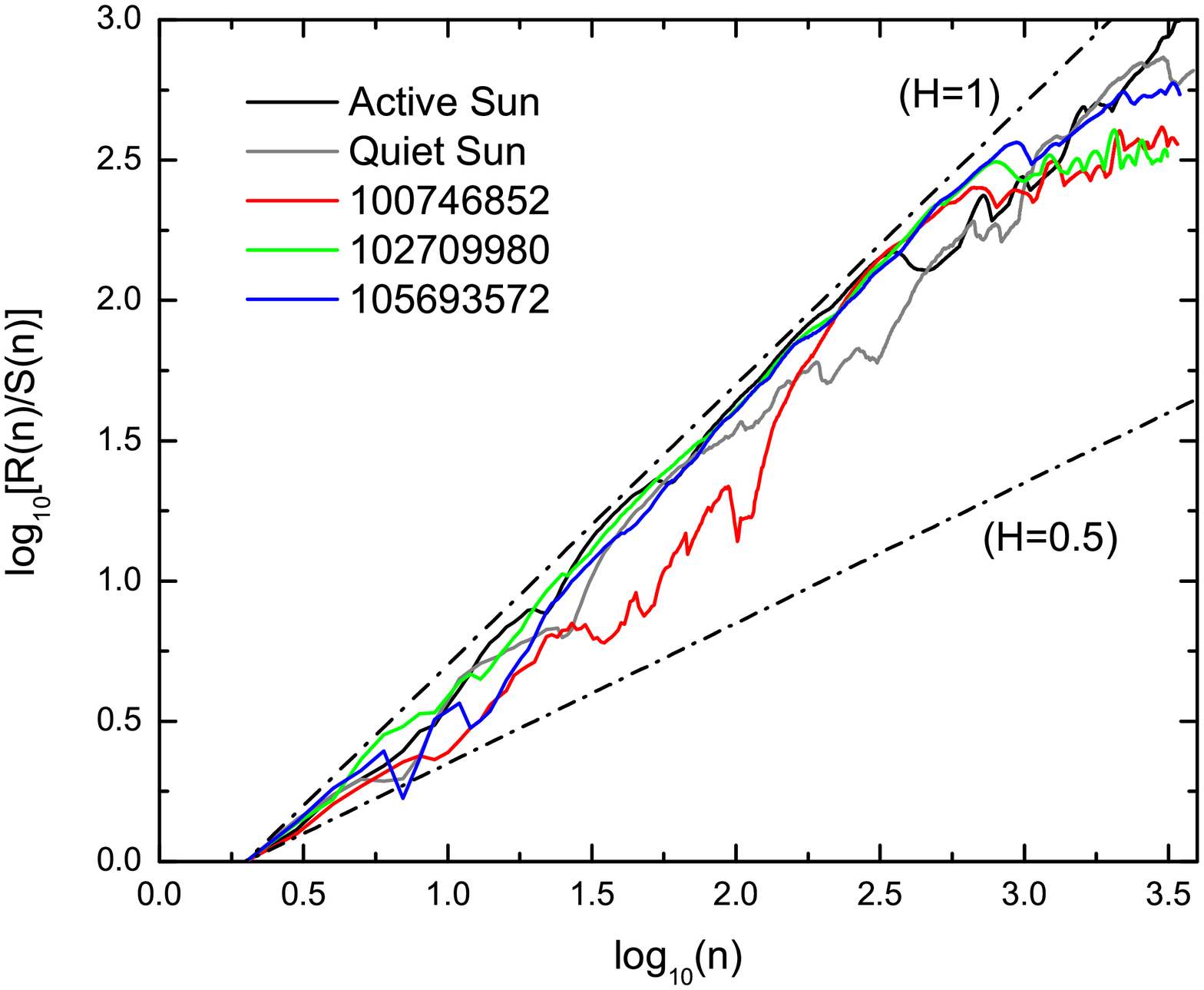}
        }%
        \subfigure[]{%
            \label{fig3b}
            \includegraphics[width=0.5\textwidth]{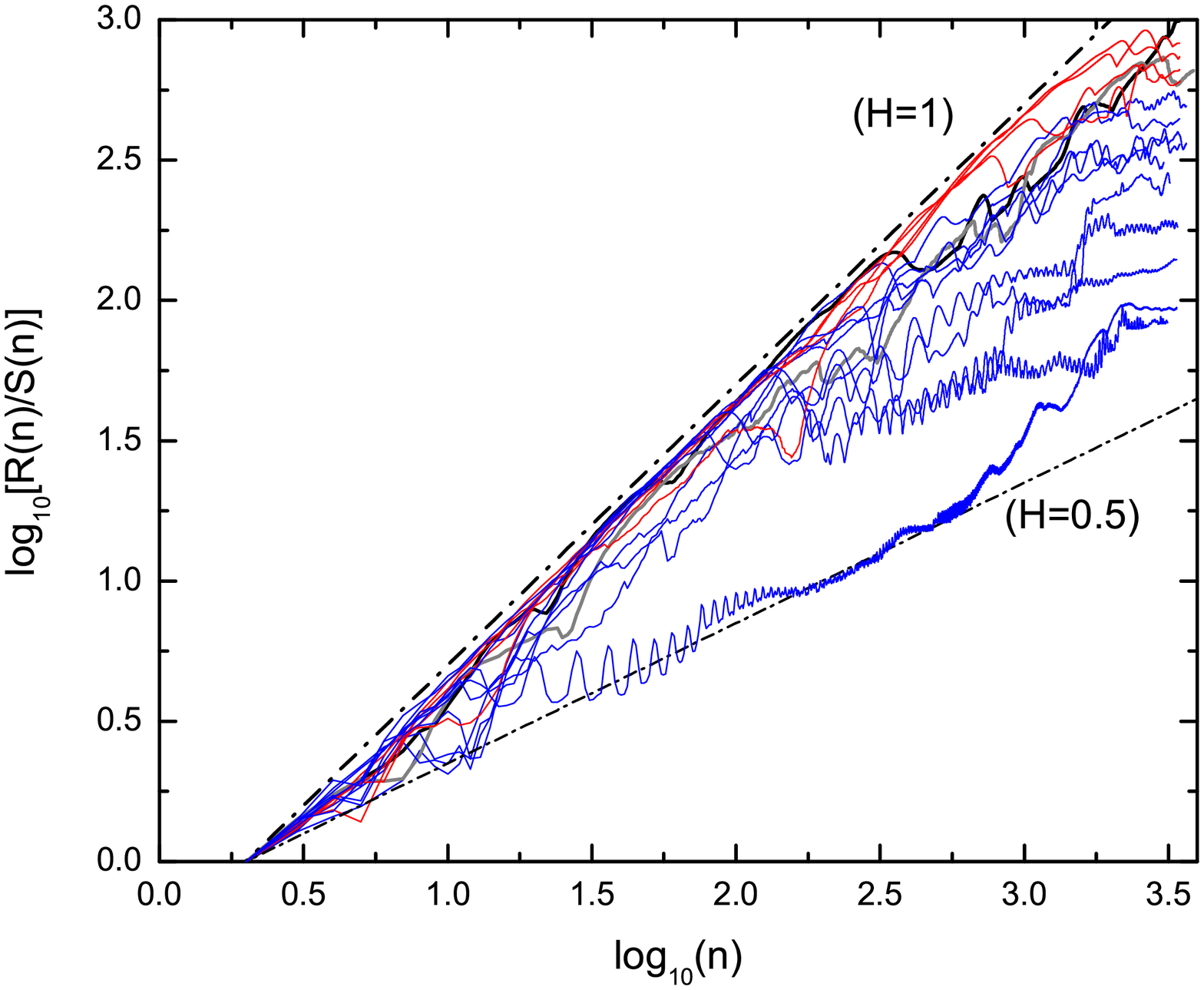}
        }\\%
        \subfigure[]{%
           \label{fig3c}
           \includegraphics[width=0.5\textwidth]{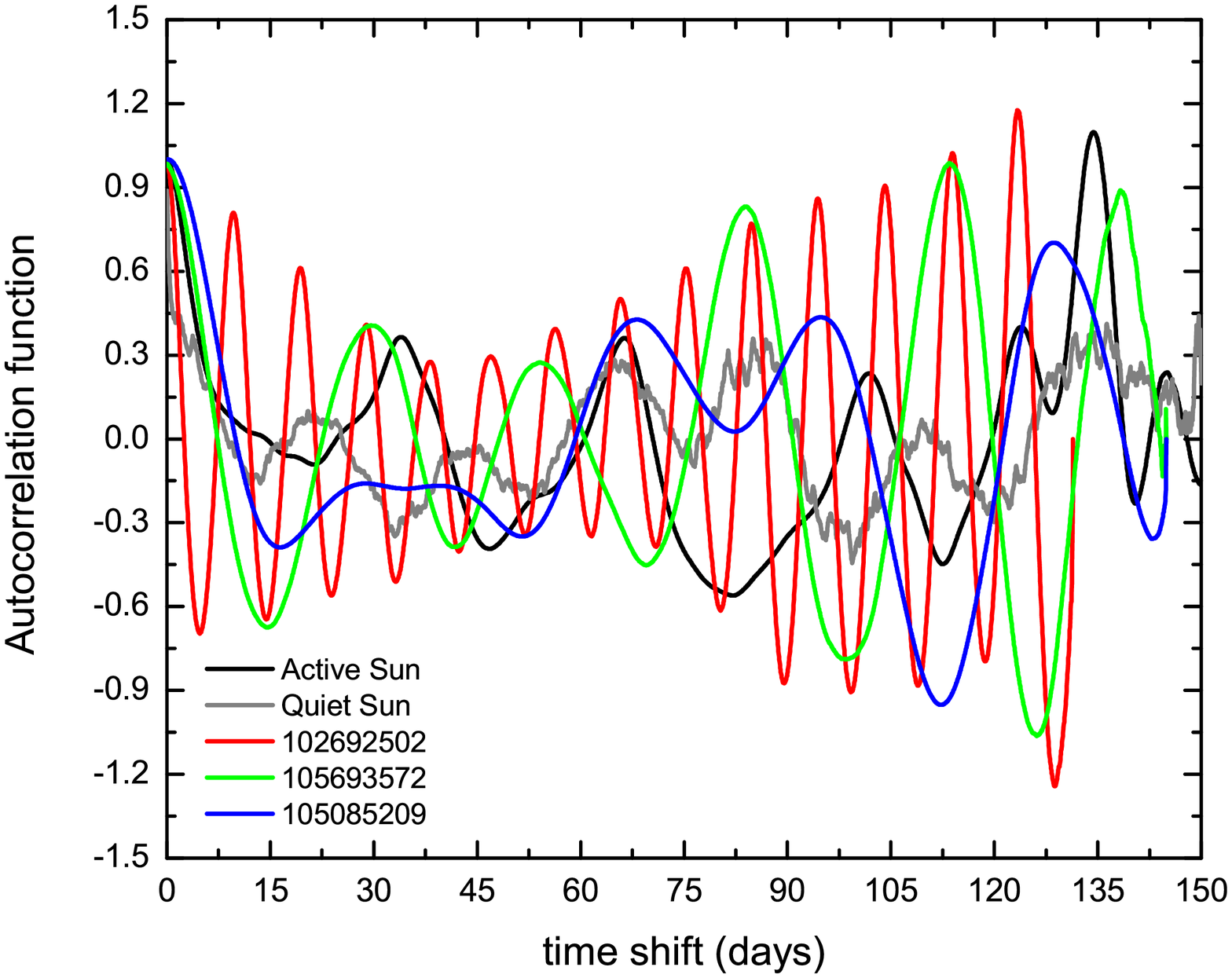}
   
        }%
       
    \end{center}
    \caption{%
        {\bf Figure (a):}  Individual $\log_{10}R(n)/S(n)$ values as a function of the box-size $n$ for the Sun (active and quiet) and ``New Sun'' candidates. Dot-dashed lines indicate the theoretically-expected behavior for purely random behavior ($H = 0.5$) and smooth sinusoidal variations ($H=1$). {\bf Figure (b):} As in the left panel, but for the CoRoT comparison stars in our sample. Red lines represent stars with periods longer than the rotation period at the solar poles, whereas the blue lines indicate those stars whose periods are less than the rotation period at the solar equator. {\bf Figure (c):} Autocorrelation function of an extract of our sample. 
     }%
   \label{fig3}
\end{figure}

The observational results described in \S\ref{sec:data}, pointing to Sun-like rotation behavior for the stars CoRoT IDs 100746852, 102709980, and 105693572, are reinforced by the behavior of the Hurst exponent, once compared with the Sun and with stars presenting non-Sun-like rotation characteristics.  Figure~\ref{fig3a} shows the values of $R/S$ for the Sun and for the ``New Sun'' candidates. In Figures~\ref{fig3a} and \ref{fig3b} it is apparent that the slope of the linear fit is affected by saturation effects in $R/S$ at high $n$ values, leading to an underestimation of the $H$ exponent in that regime. However, as can be seen from Figure~\ref{fig3b}, the linear $\log [R(n)/S(n)] - \log n$ trend otherwise clearly represents a reasonable description of the global behavior of the Hurst exponent, in the analyzed datasets. Indeed, it is clear from Figure~\ref{fig3a} that the Hurst exponent for the ``New Sun'' candidates follows the same trend as observed for the Sun itself. Our sample is a typical example where we find processes with both semi--sinusoidal cyclic components and one or more noise components. In this case, the cyclic effect creates a global statistical dependence. In particular, this effect can be exemplified by a pure sine function $f(t)=A\sin(2\pi t/Per+\phi)$ added to a stochastic noise $\epsilon(t)$, where $A$ denotes the amplitude, $Per$ the period and $\phi$ the phase (e.g.: Mandelbrot and Wallis 1969b). A method to quantify this effect is so-called AutoCorrelation Function (ACF), which may be assumed as a sine function that oscillates up and down without limit. We applied this method to our data and the result clearly reflects a semi-sinusoidal behavior as shown in Figure~\ref{fig3c}. 

By examining in detail Figures~\ref{fig3a} and \ref{fig3b}, the plot of the $R(n)/S(n)$ function presents different values of the form $n=cPer$, corresponding to sub--harmonics of period $Per$. Important details can be extracted, considering two regimes for the sub--harmonics of each period of our stellar sample. First, when $n$ is a sub--harmonic of $Per$ with $c\leq1$, the values of $R(n)/S(n)$ have no scatter or is largely reduced and, therefore, $R(n)/S(n)$ is independent of time. However, when a sub--harmonic finds ``lobes'' of decreasing amplitude, that is, when $c>1$, the effect of the scatter between subharmonics is more pronounced. This analysis shows that the gradient presents in $\log [R(n)/S(n)] - \log n$ plane for higher values of $n$ in its saturation regime, as observed in Figures~\ref{fig3a} and \ref{fig3b}, is an effect of the time window truncated at $\sim$150 days. According to Central Limit Theorem, when $n\rightarrow\infty$, each LC is characterized by a Gaussian distribution with Hurst exponent equal to $1/2$. This result shows that for LCs with finite temporal window, the $R/S$ method is significantly relevant for characterizing stars with different domains of period. This difference can be quantified by the $t$-test, from where we obtain the values 18.79 comparing the solar and super-solar regimes, 39.83 comparing the solar and sub-solar regimes and 59.35 between super- and sub-solar regimes, which corresponds to the confidence level $P<0.001$. The $t$-values were obtained by computing the average value of $R(n)/S(n)$ in each regime (including the sun) and compared two by two. 

More in detail, Figure~\ref{fig3} shows that the oscillations around the linear trendlines for our ``New Sun'' candidates are smaller than for stars with \textit{non}-sun-like rotation characteristics. As reported by Martinis et al. (2004), such an oscillating behavior suggests the presence of multiple timescale processes, related to multifractality of the analyzed LCs. Also, the wide level of variability of the Hurst exponents in our sample can be associated to the presence of deterministic chaotic dynamics with one or more centers of rotation, as evidenced by different oscillations present in Figure~\ref{fig3} (Suyal et al. 2009). This oscillating behavior reflets local nonhomogenities, remaining drifts and presence of large amplitides in LCs due to variability and lifetime of spots.

\begin{figure}
\plotone{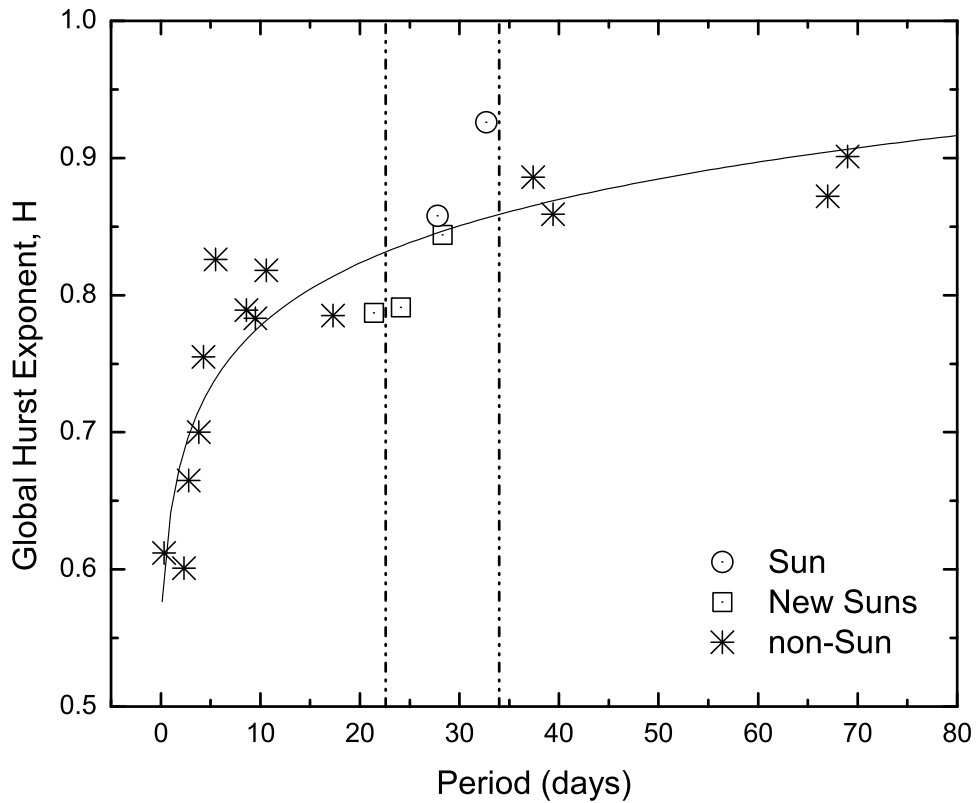}
\caption{Values of the global $H$ exponent, derived on the basis of our ``global'' $R(n)/S(n)$ analysis, as a function of rotation period, for all stars in our sample. The Sun is shown twice, since we analyzed it both in its active and quiet phases. The solid curve denotes the analytical expression that best fits the data (Eq.~\ref{per}).}
\label{fig4}       
\end{figure}

Our results reveal that the presence of these multiple timescales in stars identified as having semi-sinusoidal variability leads to a clear gradient of the ``global'' $H$ exponent, as shown in Table~\ref{tab1} and evidenced by Figure~\ref{fig4}. The latter figure displays the Hurst exponent as a function of period for all the stars of the present study (including the Sun itself, both in its active and quiet phases). From this figure, we can consider star CoRoT ID 105693572 as the best ``New Sun'' candidates because, in addition to HRD position and period, this star also has a global $H$ exponent very closely matching the solar values. More in general, our results show that, independing on the regime type, $H$ from CoRoT LCs generally exceeds $0.5$.

Finally, according to the $H$ values computed in the present study, we can tentatively define an analytical relation between the global Hurst exponent and the period as a function of time window $t$, where $t$ must be greater than three cycles (De Medeiros et al. 2013). Indeed, the data points plotted in Figure~\ref{fig4} can be nicely described by the following expression:  

\begin{equation}
\label{per}
Per=Per_{0}+\exp\left(\frac{H-H_{0}}{\sigma}\right) \quad ({\rm days}),
\end{equation}

\noindent with $Per_{0}=0.383\pm0.002$ days, $H_{0}=0.619\pm0.045$, and $\sigma=0.068\pm0.014$. This fit, obtained using a least-squares method, has $\chi^{2}/dof=0.0019$ and $R^{2}=0.777$, where $dof$ is denoted as the number of degrees of freedom and $R^2$ as the coefficient of determination of a linear regression (see Acton 1966 for futher details).

In summary, our results show that the Hurst exponent may indeed be a powerful new classifier for semi-sinousidal light curves, as clearly shown in Figure~\ref{fig4}~-- and, in particular, suggests that star CoRoT ID 105693572 is the best candidate for a ``New Sun'' to be identified so far, thus certainly well worthy of detailed follow-up studies. In addition, this star was named best canditate by presenting the set of parameters (Hurst exponent, $T_{eff}$ and Period) most similar to the Sun. Regarding rotation rates, CoRoT IDs 100746852 and 102709980 are also similar to the Sun. However, for $T_{eff}$ and Hurst exponent the results for these two stars are discrepant with the solar values. Indeed, in this work, we define as the best new Sun candidate the star with the best values for the triplet $[H,T_{eff},Per]$.

\acknowledgments
Research activity of the Observational Astronomy Stellar Board of the UFRN is supported by continuous grants from CNPq and FAPERN Brazilian agencies. We also acknowledge financial support from INCT INEspa\,{c}o/CNPq/MCT. We thanks CoRoT Teams for successful conduction of this mission. Support for M. Catelan is provided by the Chilean Ministry for the Economy, Development, and Tourism's Programa Iniciativa Cient\'{i}fica Milenio through grant P07-021-F. We thank warmly the anonymous referee for careful reading and suggestions that largely improved this work.

\begin{deluxetable}{lccccccc}
\tabletypesize{\scriptsize}
\tablecaption{Parameters of our stellar sample and the values of the global Hurst exponent $H$.}
\tablewidth{0pt}
\tablehead{{\bf Star}\tablenotemark{(a)} & RA(J2000) & DEC & Run\tablenotemark{(b)} &
$T_{\textrm{eff}}$\tablenotemark{(c)} & 
Per & $\delta Per$ & $H$\\
 &  &  &   & (K) & (days) &(days) & 
}
\startdata
\multicolumn{8}{c}{ ``New Sun'' candidates as Compared with the Sun}\\
\hline
{\bf Sun}\tablenotemark{(d)}	&		&	&	Virgo/SoHO	&		5772	&	32.7 & 0.4	& 0.926 (Active Sun)\\ 
	&		&		&		&		&	27.8 & 0.3 &	0.858 (Quiet Sun)\\ 
{\bf ID  100746852 }	&	291.094	&	1.343	&	LRc01	&		5835	&	24.1007& 0.3 &	0.791\\ 
{\bf ID  102709980 }	&	100.963	&	-0.651	&	LRa01		&		5908	&	21.3669& 0.1	& 0.787\\ 
{\bf ID  105693572 }	&	280.479	&	7.987	&	LRc02		&		5761	&	28.2595& 0.5	& 0.844\\ 
\hline
\multicolumn{8}{c}{Comparison Stars}\\
\hline
\multicolumn{8}{c}{Super-Solar Sample}\\
\hline
{\bf ID 105085209}	&	279.601	&	7.646	&	LRc02		&		4179	&		39.4494 & 0.9	& 0.859\\
{\bf ID 105284610}	&	279.873	&	7.679	&	LRc02		&		4421	&		37.3701 & 0.4	& 0.886\\
{\bf ID 105367925}	&	279.991	&	6.731	&	LRc02		&		4842	&		68.9516 & 1	&  0.901\\
{\bf ID 105379106}	&	280.006	&	7.311	&	LRc02		&		4401	&		67.2065 & 1	& 0.872\\
\hline
\multicolumn{8}{c}{Sub-Solar Sample}\\
\hline
{\bf ID 101121348}	&	291.644	&	-0.053	&	LRc01		&		5989	&		17.3479& 0.2	& 0.785\\
{\bf ID 101710670}	&	292.671	&	-0.023	&	LRc01		&		6082	&		5.4984& 0.02	& 0.823\\
{\bf ID 102692502}	&	100.873	&	0.001	&	LRa01		&		15526	&		9.5111 & 0.02	& 0.783\\
{\bf ID 102752622}	&	101.208	&	-1.065	&	IRa01/LRa01		&		6558	&		2.3321 & 0.0003	& 0.601\\
{\bf ID 102770893}	&	101.311	&	-1.217	&	IRa01/LRa01		&		6140	&		4.2850 & 0.002	& 0.755\\
{\bf ID 105503339}	&	280.181	&	7.651	&	LRc02		&		4253	&		10.600 & 0.4	& 0.818\\
{\bf ID 105665211}	&	280.436	&	5.416	&	LRc02		 	&		8388	&		3.8148 & 0.003	& 0.700\\
{\bf ID 105945509}	&	280.908	&	5.744	&	LRc02		&		4725	&		2.7917 & 0.003	& 0.665\\
{\bf ID 105957346}	&	280.933	&	6.441	&	LRc02	&		4236	&		0.3341 & 5$\cdot10^{-5}$ &	0.612\\
{\bf ID 105845539 }	&	280.705	&	7.556	&	LRc02		&		5809	&		8.610	& 0.4 & 0.789\\ 
 \enddata
\tablenotetext{(a)}{Following the usual CoRoT nomenclature (see, e.g., Baglin 2006).}
\tablenotetext{(b)}{CoRoT IDs are given for all stars, except the Sun.}
\tablenotetext{(c)}{From Sarro et al. (2013).}
\tablenotetext{(d)}{Quoted values are from \citet{lanza2003} and \citet{em12}.}
\label{tab1}
\end{deluxetable}

\end{document}